\documentstyle[pra,aps,floats,twocolumn,graphicx]{revtex}
\newcommand{\position}{tbp}

\newcommand{\ket}[1]{\mbox{$\left|#1\right\rangle$}}
\newcommand{\bra}[1]{\mbox{$\left\langle#1\right|$}}

\newcommand{\ex}{\mbox{\bf e$_x$}}

\newcommand{\vect}[1]{\mbox{\bf #1}}
\newcommand{\DO}{\Delta\omega}

\newcommand{\sub}[1]{_{\text{#1}}}

\newcommand{\stau}{s_\tau}

\newcommand{\GPcm}{\frac{\text{G}}{\text{cm}}}

\title{Trapped-Atom-Interferometer in a Magnetic Microtrap}

\author{W. H\"ansel\footnote{Corresponding author:\\Fax: ++49-89/285192, E-mail: Wolfang.Haensel@mpq.mpg.de}, J. Reichel, P. Hommelhoff, and T.W. H\"ansch}

\address{Max-Planck-Institut f\"ur Quantenoptik and Sektion Physik
der
  Ludwig-Maximilians-Universit\"at\\Schellingstr. 4, D-80799
  M\"unchen, Germany}

\draft

\date{\today}

\begin{document}

\maketitle

\begin{abstract}
We propose a configuration of a magnetic microtrap which can be
used as an interferometer for three-dimensionally trapped atoms.
The interferometer is realized via a dynamic splitting potential
that transforms from a single well into two separate wells and
back. The ports of the interferometer are neighboring vibrational
states in the single well potential. We present a one-dimensional
model of this interferometer and compute the probability of
unwanted vibrational excitations for a realistic magnetic
potential. We optimize the speed of the splitting process in order
suppress these excitations and conclude that such interferometer
device should be feasible with currently available microtrap
technique.
\end{abstract}

\pacs{03.75.-b, 03.65.-w, 39.20.+q, 39.25.+k, 39.90.+d}

\section{Introduction}
\label{section:Introduction}

Since the first realization of magnetic traps
\cite{Reichel99,Folman2000} and guides \cite{Mueller99,Dekker2000}
with current-carrying conductors on a chip, a large variety of
magnetic potentials have become experimentally accessible, which
would be impractical or even impossible to realize with
macroscopic coils. 
The splitting of two-dimensionally trapped atom clouds has been
demonstrated \cite{Mueller2000,Cassettari2000}, and recently, we
were able to split and unite a three-dimensionally trapped cloud
of rubidium atoms in a chip trap \cite{Haensel2001}.

Current experiments aim at populating single quantum states of
such microtrap potentials with either an atomic ensemble (i.e.,
creating a Bose-Einstein condensate), or indeed with a single
atom.  One promising application of such a system would be an
integrated atom interferometer on a chip \cite{Hinds2001}. The
small size and monolithic construction of such a device suggests
its suitability for ``real-word'' applications. Moreover, the fact
that magnetic potentials may be ``engineered'' on the chip enables
novel interferometer schemes with features quite different from
more traditional atom interferometers [...]. Here we study a
scheme in which the particle wave of a single, trapped atom is
coherently split up and reunited by a time-varying magnetic
potential (fig.\,1). Splitting occurs in one dimension, while
tight confinement in the remaining two dimensions leads to an
effective 1D situation. This is in contrast to \cite{Hinds2001}
where the dynamics of the splitting is in two dimensions. As
depicted in fig.\,1, interference occurs between the lowest two
vibrational states, \ket{\varphi_0} and \ket{\varphi_1}, of the
splitting potential (the internal atomic state remains unchanged).
A phase-changing interaction in one ``arm'' (stage II in fig.\,1)
translates into a change of the relative populations in
\ket{\varphi_0} and \ket{\varphi_1} when the potential is
recombined. As in other interferometers, a longer duration of
stage II leads to a larger accumulated phase (i.e., a larger arm
length). However, unlike the situation in most free-atom schemes
and the guided-atom scheme proposed in \cite{Hinds2001}, in our
scheme the atom does not move nor does its wave function spread
during this stage: the propagation along the traditional
interferometer path is replaced by the evolution in a constant
potential (stage II), which leaves the position and the physical
size of the wave function unchanged. This interferometer is thus
particularly well suited to measure local fields and interactions,
which presents an advantage over experiments with propagating
atoms\footnote{There is a subtle difference between atom
interferometers with beams and with trapped atoms: in spatial beam
splitters atoms are slowed down when the energy of the transverse
state increases. There are currently studies on the way how this
effect can be explored for enhanced detection schemes of the
outgoing state\cite{Andersson2001}.}. One could, for instance,
measure the phase shift arising from a two-body collision
\cite{Calarco99} or the amount of decoherence induced from a
nearby surface \cite{Wilkens99B}.

 \begin{figure}[\position]
 \center
 \includegraphics{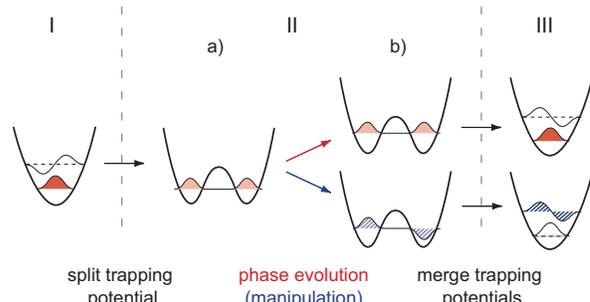} \caption{Scheme of the trapped atom interferometer:
 one or several atoms are prepared in the vibrational ground state of the single well potential
 (I). When the well separates, the wave function evolves
 adiabatically into a symmetric delocalized state (II a). The phase of the wave function in each potential well
 can be assumed to evolve independently and monitors sensitively external electric and magnetic field gradients (II b).
 As the potential wells reunite, the antisymmetric state transforms into
 the first exited vibrational state whereas the symmetric one retransforms into the ground state (III).}
 \label{fig:Interferometer}
 \end{figure}

In this paper, we present a detailed analysis of this
interferometer scheme, employing a realisitic magnetic potential
which can be implemented with currently available microtrap
technique. We consider the case of an individual trapped atom, a
situation which is also targeted by experiments under way (for a
study of a BEC in an idealized 1D potential, see
\cite{Menotti2001}). The potential is created by the simple
conductor configuration shown in fig.\,2. The current $I_0$
together with the external field $B_{0,y}$ provides tight
confinement in the $yz$-plane. The two currents $I\sub{ext}$
together with the homogeneous field component $B_{0,x}$ close this
2D trap in axial direction, completing the single-well potential
(fig.\,\ref{fig:WaveFunctions}\,a, left). The current $I_c$
creates an adjustable ``bump'' in the center of this trap, and
thus induces the splitting. Increasing $I_c$ transforms the
potential from single-well to double-well
(fig.\,\ref{fig:WaveFunctions}\,a, right), in loose analogy with
the first passage through the beam-splitter of a Michelson-Moreley
interferometer for light.

To achieve a good fringe contrast, it is essential that no
higher-lying vibrational states be excited during this splitting.
Therefore, the crucial part of the interferometer is the quantum
dynamics during the splitting and merging process. The splitting
(merging) of the wave functions occurs as the quantum states
adiabatically evolve in the varying potential. We analyze these
dynamics in a one-dimensional model, using analytic expressions
for the microtrap magnetic field.
We numerically determine the energy eigenstates of $^{87}$Rb atoms
in the given potential and then trace the dynamics of an initial
state, using the eigenstates as a time-dependent basis.

We show that successive vibrational levels in the initial trap
evolve into pairs of degenerate states when the
potential is split. In this ``sensing state'', the wave functions
are composed of two identical oscillator states in the left and in
the right well. Either of the two parts can acquire a phase shift
independent of the other one, reflecting e.g. an additional small
field gradient or the presence of an additional atom in one of the
wells. When the potential is transformed back into a single well,
the population of the vibrational levels depends on the phase
difference that is picked up in the degenerate states.
Fig.\,\ref{fig:Interferometer} illustrates this process: first,
the system (i.e. one or several atoms) is prepared in the
vibrational ground state. Upon separation of the potential, this
state evolves into a symmetric state that spreads over the two
potential wells. In an analogous manner, the antisymmetric first
vibrational level transforms into an antisymmetric delocalized
state. As the system's Hamilton operator is symmetric throughout
the whole process, it cannot induce transitions between states of
opposite symmetry, and the eigenstates can always be chosen of
well-defined parity.

If the symmetric and antisymmetric state are spatially separated
far enough, they degenerate, and the left (right) localized state
can be constructed as sum (difference) of the symmetric and
antisymmetric state. A perturbation of the potential, which does
not have even parity, will lift this degeneracy in favour of the
localized states. These localized states make up for the classical
interferometer arms, measuring very sensitively deviations from an
ideal symmetric potential or interactions with other atoms.

In the following section, we investigate the separation process
using the 1D-potential taken from the microtrap device sketched in
fig.\,\ref{fig:Layout}. We establish the quantum mechanical
equation of motion and use first order perturbation theory to
determine the amount of vibrational excitations. Assuming a linear
variation of the current $I_c$, we find the excitation probability
lower than 2\% if the separation takes 60\,ms or longer. This
indicates that an experimental realization should be possible, and
the situation can still be improved when an arbitrary variation of
$I_c$ is allowed. We therefore dedicate section
\ref{section:Optimization} to a method which minimizes
non-adiabatic excitations by finding the most appropriate time
dependence for the shape of the potential (here controlled via
$I_c$). Such method is of interest not only for interferometers.
It applies to all cases of time-dependent potentials, and it can
even be transferred to spatially varying potentials such as
beam-splitters. For our interferometer, this method helps to
reduce the splitting time by a factor of two, at the same time
reducing the excitation probability by more than a factor ten.

\section{The Trapped Atom Interferometer}
\label{section:MainPart} 
\begin{figure}[\position]
 \center
 \includegraphics{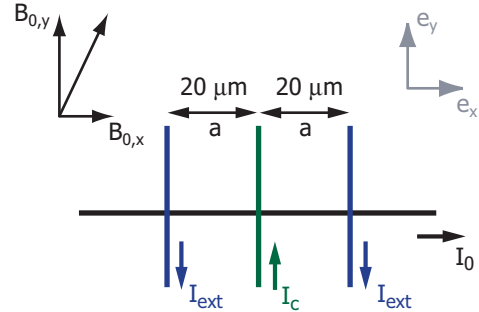} \caption{Layout of the interferometer conductor pattern.}
 \label{fig:Layout}
 \end{figure}

The microtrap device that we propose for
the interferometer is a symmetric arrangement of wires as depicted
in fig.\,\ref{fig:Layout}. Its potential is similar to the one
that we used in the merging experiment with thermal atoms
\cite{Haensel2001}, but it is scaled down to a wire distance of
20$\,\mu m$ and simplified to produce a strictly symmetric
potential. The quantum state computations are made for $^{87}$Rb
atoms in the $\ket{F=2,m_F=2}$ ground state, the effective
potential being $U(x) \approx h\cdot 1.4\,\mbox{MHz}\cdot \vect
B(x)/\mbox{G}$.

The current $I_0$ in the central wire and the homogeneous field
$B_{0,y}$ create a two dimensional quadrupole field which strongly
confines the atoms in the $yz$-plane. Each of the crossing wires
contributes a longitudinal field modulation of Lorentzian shape
(see \cite{Reichel2000}):
\begin{figure}[\position]
 \center
 \includegraphics[height=6.9cm]{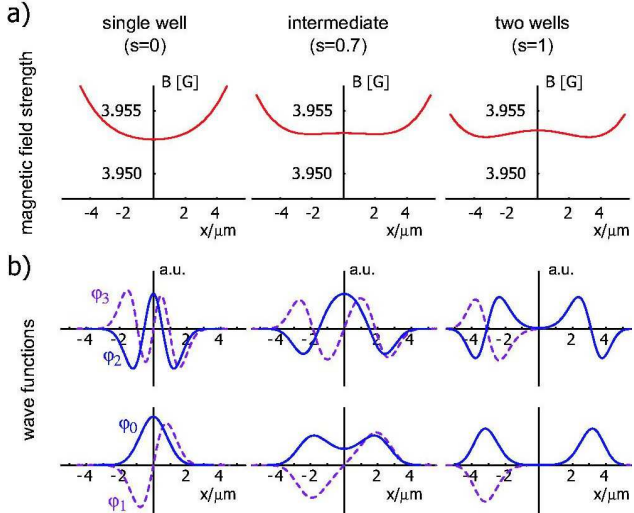} \caption{a) Shape of the
 magnetic splitting potential for characteristic values of the control
 parameter $s$ (see eq.~(\ref{eq:IExt}) and (\ref{eq:IC})\,),
 b) eigenstates of $^{87}$Rb atoms in the specified potential.
 }
 \label{fig:WaveFunctions}
 \end{figure}
 \begin{figure}[\position]
 \center
 \includegraphics[height=5.1cm]{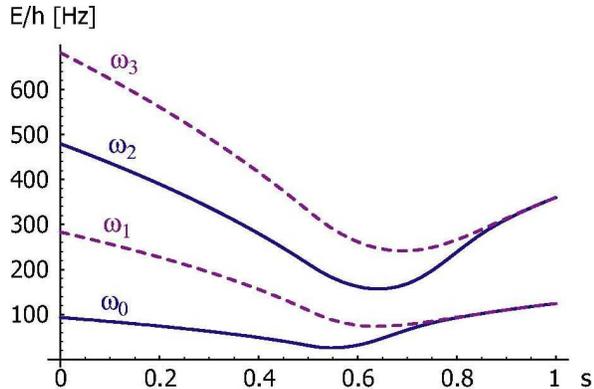} \caption{Energy eigenvalues of the
 system's Hamilton operator
 as the trapping potential is divided into two wells. Neighboring states
 of opposite symmetry form pairs and degenerate as the potential wells
 separate.}
 \label{fig:Eigenvalues}
 \end{figure}
The two currents $I\sub{ext}$ together with the field component
$B_{0,x}$ generate two valleys along the longitudinal axis, which
do not appear separate if the trap is located far enough from the
surface. The current $I_c$ with its direction opposite to the two
external currents is used to split the Ioffe-Pritchard potential
into two neighboring wells (fig.\,\ref{fig:WaveFunctions}\,a).
Choosing the parameters as

 \begin{eqnarray}
 I_0      & = & 525\,\mbox{mA} \\
 B_{0,y}  & = & 20\,\mbox{G} \\
 B_{0,x}  & = & 16\,\mbox{G} \\
 I_{ext}  & = & 140\,\mbox{mA} + 2.91\,\mbox{mA}\cdot s \label{eq:IExt}\\
 I_{c}    & = & 0.25\,\mbox{mA} + 4.4\,\mbox{mA}\cdot s \label{eq:IC} \mbox{ ,}
 \end{eqnarray}
the trap is located $35\,\mu$m above the surface, yielding a
transversal oscillation frequency of $\omega\sub{trans}\approx
2\pi\,53.7$\,kHz. The parameter $s$ determines the shape of the
trap, running from 0 for one single well to 1 for separated wells.
The point $s=1$ has been chosen such that the two lowest
vibrational levels of each well are clearly separated (i.e., the
two lowest sets of states are both degenerate). The
time-dependence of the system's Hamiltonian is expressed via the
function $s(t)$. In a simple approach, $s$ may be chosen to vary
linearly in time, but as we will discuss in
section~\ref{section:Optimization}, an optimized function $s(t)$
can be found which minimizes vibrational excitations during the
splitting (merging) process.

In fig.\,\ref{fig:WaveFunctions}\,a, the resulting magnetic field
along the longitudinal axis \ex\ is displayed for characteristic
values of $s$; the transverse potential minimum is plotted against
the longitudinal position. The plots below show the eigenstates of
$^{87}$Rb atoms (\ket{F=2,m_F=2}) in this field as they are
numerically computed from the Schroedinger equation
(eq.~(\ref{eq:Schroedinger}) below). The corresponding energy
eigenvalues, measured relative to the minimum value of the
potential, are given in fig.\,\ref{fig:Eigenvalues}.

For $s=0$, the four lowest levels correspond to the states of a
harmonic oscillator with quantum number $n=0,\ldots,3$ and
oscillation frequency \mbox{$\omega_{s=0} \approx 2\pi\,190$\,Hz}.
We use the quantum number $n$ to identify the eigenstates as
$\ket{\varphi_n(s)}$ throughout the whole evolution.  As the
value of $s$ is raised, the vibrational levels 
evolve into symmetric and antisymmetric delocalized states. At
$s=1$, the four lowest levels
form two sets of degenerate states, their energy being
$E_{2k}=E_{2k+1}=\hbar\omega_{s=1}(k+\frac12)$,
$k=\left\{0,1\right\}$, $\omega_{s=1} \approx 2\pi\,240$\,Hz. At
each stage, the separation of the transverse levels ($>$ 50\,kHz)
is much larger than the separation of longitudinal states
involved. For this reason, the longitudinal states do not
intermingle with the transverse levels, even if the system's
symmetry is slightly disturbed. The quantum dynamics is therefore
adequately described by a one-dimensional model.

In order to make the interferometer work properly, the atomic wave
function should follow ideally the (time-dependent) eigenstates
$\ket{\varphi_k(t)}$ of the system. If the potential is varied too
fast, the evolution is non-adiabatic, i.e. vibrational excitations
are generated. For the investigation of these excitations, we will
focus on the first half of the interferometer cycle: we use a
time-dependent interaction picture to compute the time scale on
which the separation process can be lead adiabatically.

The (time-dependent) basis for the computation is found by solving
the time-independent Schr\"odinger equation
 \begin{equation}
 \hat{H}(s)\ket{\varphi_k(s)}  =  \hbar \omega_k(s)\,\ket{\varphi_k(s)}
 \label{eq:Schroedinger}
 \end{equation}
with the Hamilton operator
 \begin{equation}
 \hat{H}(s)=\frac{\hat{\vect{p}}^2}{2m} +
 \mu_B\,g_F\,m_F\,\left|{\vect{B}(s,\hat{\vect{r}})}\right|
 \label{eq:Hamiltonian}
 \mbox{\,,}
 \end{equation}
where $s$
takes the role of a mere parameter. For the given magnetic field,
the eigenfunctions have been computed numerically and are
displayed in fig.\,\ref{fig:WaveFunctions}\,b.

The natural phase evolution of the eigenstates can be included
into the basis and yields the ansatz
 \begin{equation}
 \label{eq:ansatz}
 \ket{\psi(t)} = \sum_k c_k(t)\,e^{-i\int_0^{t'} \omega_k(t')\,dt'}
 \ket{\varphi_k(t)} \mbox{ .}
 \end{equation}
The equation of motion for the coefficients $c_k(t)$ is obtained
when eq.~(\ref{eq:ansatz}) is inserted in the time-dependent
Schr\"odinger-equation with the Hamiltonian
(\ref{eq:Hamiltonian})\footnote{The time-dependence of
$\omega_k(t)$ and $\ket{\varphi_k(t)}$ is explicit through the
control parameter $s$: $\omega_k(t)\equiv\omega_k(s(t))$ etc.}:
 \begin{eqnarray}
 \frac{d}{dt}c_k(t) &= -\sum_n &c_n(t)\,e^{i
 \int_0^t\left(\omega_k(t')-\omega_n(t')\right)dt'} \nonumber\\
 &&\;\cdot\bra{\varphi_k(t)}\frac{d}{dt}\ket{\varphi_n(t)}
 \mbox{ .}
 \end{eqnarray}
Given that a single eigenstate $\ket{\varphi_i(t=0)}$ is prepared
in the beginning, and further assuming that the transition
probability into other vibrational states is small, first order
perturbation theory can be used to determine the coefficients
$c_f$ and the corresponding transition probabilities $P_{if}$:
 \begin{eqnarray}
 c_f(t) &=& \int_0^t e^{i
 \int_0^t\left(\omega_f(t')-\omega_i(t')\right)dt'}
 \bra{\varphi_f(t)}\frac{d}{dt}\ket{\varphi_i(t)}\,dt
 \label{eq:TransitionAmplitude} \\
 P_{if}(t)&=&\left| c_f(t) \right|^2\mbox{ .}
 \label{eq:TransitionProbability}
 \end{eqnarray}

The coupling 
$\bra{\varphi_f(t)}\frac{d}{dt}\ket{\varphi_i(t)}=
\bra{\varphi_f(s)}\frac{d}{ds}\ket{\varphi_i(s)}\,\frac{ds}{dt}$
to higher levels is directly proportional to the rate
$\frac{ds}{dt}$ at which the control parameter $s$ is changed.
Therefore, if all levels are separated by a minimum energy
$\hbar\omega_0$, the transition amplitudes can be made negligible
by choosing an appropriate duration for the process. Conversely,
if at certain instants some energy levels degenerate, this will
create large transition amplitudes unless the coupling coefficient
$\bra{\varphi_f(s)}\frac{d}{ds}\ket{\varphi_i(s)}$ between these
levels vanishes at the points of degeneration. In the trapped atom
interferometer presented here, we encounter such degenerate
levels. But as the states that degenerate are of opposite symmetry
throughout the complete evolution, the coupling coefficient
remains zero for all times. Therefore, the excitation probability
can be made arbitrarily small by choosing the process duration
long enough.

\begin{figure}[\position]
 \center
 \includegraphics[height=5.1cm]{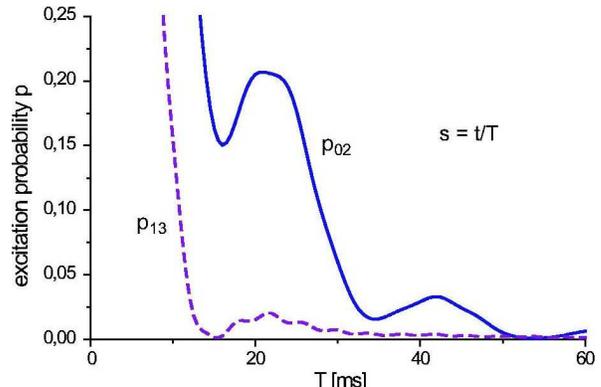} \caption{Excitation probability
 for the transitions
 $\ket{\varphi_0}\rightarrow\ket{\varphi_2}$ and
 $\ket{\varphi_1}\rightarrow\ket{\varphi_3}$
 for a linear increase of the separation $s$ with time.}
 \label{fig:ExcitationI}
 \end{figure}

This consideration is confirmed by numerically evaluating
expressions~(\ref{eq:TransitionAmplitude}) and
(\ref{eq:TransitionProbability}) for either of the interferometer
levels $\left( \ket{\varphi_0},\ket{\varphi_1} \right)$. In a
first approach, the separation parameter has been chosen linear in
time $s=t/T$. Fig.\,\ref{fig:ExcitationI} shows the transition
probabilities into the neighboring interferometer levels which
contribute largest to all vibrational excitations. The data
indicate that the excitation probability is less than 1\% if the
separation process takes longer than 60\,ms.

This is an encouraging result, as it seems experimentally
realizable. Moreover, as the time-dependence of the potential can
be freely chosen, one can adjust the speed of the separation
process $\frac{ds}{dt}$ in order to further reduce excitations.
This is of general interest, because a linear variation of the
control parameter $s$ ís not necessarily the best choice. Indeed,
one wishes to find a method that optimizes the process
irrespective of its parametrization.

\section{Optimization}
\label{section:Optimization}

In this section, we develop a scheme which minimizes vibrational
excitations in time-dependent potentials. In a slight variant,
this method can equally be used to find an adequate shape for a
beam-splitter potential.

In order to optimize the adiabaticity of the separation process we
first take a look at the coupling term from eq.
(\ref{eq:TransitionAmplitude})
 \begin{equation}
 \bra{\varphi_f(t)}\frac{d}{dt}\ket{\varphi_i(t)}=
 \bra{\varphi_f(s)}\frac{d}{ds}\ket{\varphi_i(s)}\,\frac{ds}{dt}\mbox{
 ,}
 \end{equation}
which is proportional to the process speed $\frac{ds}{dt}$ and to
the coupling coefficient
$a(s)\equiv\bra{\varphi_f(s)}\frac{d}{ds}\ket{\varphi_i(s)}$.

Intuitively, one can increase the process speed $\frac{ds}{dt}$ if
$a(s)$ is small, and decrease it in the opposite case.
Furthermore, the process speed should be adapted to the energy
difference of the levels involved, being the more increased the
further the energy levels lay apart from each other. Last not
least, one has to avoid discontinuities in the process speed
including the start and the end of the separation. In the
following, these intuitive rules will be substantiated into a set
of differential equations to yield an optimized process control
$s(t)$.

We assume that the process is lead during $0 \leq t \leq T$ and
that the separation parameter at $t=T$ is $s(t\!=\!T)=1$. Indeed,
we want to fix a shape of the control parameter $s$ which does not
depend on the process duration. Therefore, we implicitly assume
that $s(t) \equiv s(t,T)$ can be written as
 \begin{equation}
 s(t,T)=s(\frac t T,1)\,.
 \end{equation}
The goal is then to fix some maximum excitation probability
$\epsilon^2$ and to find an appropriate shape for the function
$s(t,T)$ which minimizes $T\sub{adiab}$ fulfilling the condition
 \begin{equation}
 \left|c_f(T)\right| \leq \epsilon \;\;\; \forall T \geq
 T\sub{adiab}\mbox{\,.}
 \end{equation}
If, by some chance,
the distance of energy levels
$\Delta\omega\left(s(t')\right)\equiv \omega_f - \omega_i$ is
constant throughout the process, the transition amplitude $c_f(T)$
appears as the Fourier transform of $a(s)\,\frac{ds}{dt}$:
 \begin{eqnarray}
 c_f(T) &=& \int_0^T e^{i\int_0^t\DO(t')\,dt'}a\left(s(t)\right)\frac{ds}{dt}\,dt \label{eq:AmplitudeExplicit} \\
        &=& \int_0^T e^{i\DO\,t}a(s)\frac{ds}{dt}\,dt \hspace{1cm}\mbox{ for $\DO=const.$}
 \end{eqnarray}
If, in addition, $a(s)$ happens to be constant over the process,
the solution of the problem is simple: the shape of the process
speed $\frac{ds}{dt}$ should be chosen such that it produces the
least amount of side bands possible in a Fourier transformation.
An appropriate shape would e.g.\ be a Blackman pulse
\cite{Blackman58}:
 \begin{equation}
 \frac{ds}{dt} = \frac 1 T \left(1-\frac{25}{21}\cos(2\pi \frac t T)+\frac{4}{21}\cos(4\pi \frac t T)
 \right)\,,
 \end{equation}
which can be directly integrated to yield $s(t)$.

The idea of the Fourier transform can be extended to the more
general case. A substitution of the time variable $t$ by some new
variable $\tau$ can be made in a way that the argument of the
exponential in eq.~(\ref{eq:AmplitudeExplicit}) becomes linear in
$\tau$
 \begin{equation}
 \int_0^t\DO(t')\,dt' = \frac{T}{T_0}\,\tau(t)
 \label{eq:Substitution}
 \end{equation}
and that $\tau$ runs from 0 to 1 during the process.
The time scale $T_0$ will be part of the optimization result.
Equation (\ref{eq:AmplitudeExplicit}) then assumes the form of a
Fourier transform of some new expression $u(\tau)$~\footnote{In
the following equations, the index $\tau$ marks the fact that the
functional dependence of the parameter $s$ is on $\tau$, not on
$t$.}:
 \begin{eqnarray}
 c_f(T) &=& \int_{0}^{1} e^{i\frac{T}{T_0} \tau}u(\tau)\,d\tau \label{eq:AmplitudeSimplified}\\
 u(\tau)&\equiv& a(s)\frac{d\stau}{d\tau} \label{eq:UTau}\mbox{ .}
 \end{eqnarray}

The expression $u(\tau)$ is a generalized coupling term, acting in
the transformed time frame $\tau$. As above, one can now choose a
shape for this coupling term $u(\tau)$ (however, not its
amplitude) and will obtain the probability amplitude as its
Fourier transform.

After the optimization strategy is chosen, it remains to solve the
equations~(\ref{eq:Substitution}) and (\ref{eq:UTau}). One might
be tempted to deduce the relation $\frac{dt}{d\tau}$ from
eq.~(\ref{eq:Substitution}) and insert it into eq.~(\ref{eq:UTau})
to solve directly for $s(t)$. Unfortunately, this results in an
intractable problem. Instead, one can take advantage of the
substitution already made and first solve for $\stau(\tau)$. The
relation between $\tau$ and $t$ is then established in a second
step. This way, the problem is split into two differential
equations the first of which gives the amplitude of $u(\tau)$, and
the second of which determines the time scale $T_0$ used in the
substitution. These two values determine size and scale of the
probability amplitude $c_f(T)$.

The first differential equation involves the shape of the function
that is chosen for the generalized coupling term $u(\tau)$, and it
is a direct consequence of eq.~(\ref{eq:UTau}):
 \begin{equation}
 \frac{d\stau}{d\tau} = \frac{u(\tau)}{a(\stau(\tau))} \label{eq:DEQ1} \mbox{ .}
 \end{equation}
It is important to note, that although the time
$\frac{d\stau}{d\tau}$ is used to shape the coupling term
$u(\tau)$, its amplitude does not correspond to the overall
process speed. Instead, the amplitude of $u(\tau)$ has to be
adjusted such that the solution matches the boundary conditions
$s(\tau\!=\!0)=0$, $s(\tau\!=\!1)=1$. This can for instance be
done by iteratively solving eq.~(\ref{eq:DEQ1}) for different
amplitudes of $u(\tau)$.

The second differential equation establishes the relation between
$t$ and $\tau$ and arises from the substitution of $t$
(eq.~(\ref{eq:Substitution})\,), once that $\stau(\tau)$ has been
determined:
 \begin{equation}
 \frac{d\tau}{dt} = \DO\left(\stau(\tau)\right)\frac{T}{T_0} \mbox{ .}
 \end{equation}
Choosing $T=T_0$, this equation can be solved numerically, and one
finds $T_0$ as the point in time, for which $\tau(t)$ reaches its
boundary $\tau(T_0)=1$.

The result for the transition amplitude is now completely
described by equation~(\ref{eq:AmplitudeSimplified}), the
amplitude of $u(\tau)$ and the time scale resulting from the
choice of the pulse shape. The optimized evolution of the control
parameter is computed from the concatenation of $s_\tau(\tau)$ and
$\tau(t)$:
 \begin{equation}
 s(t) = \stau\bigl(\tau(t\frac{T_0}{T})\bigr) \mbox{ .}
 \end{equation}

 \begin{figure}[\position]
 \center
 \includegraphics[height=5.3cm]{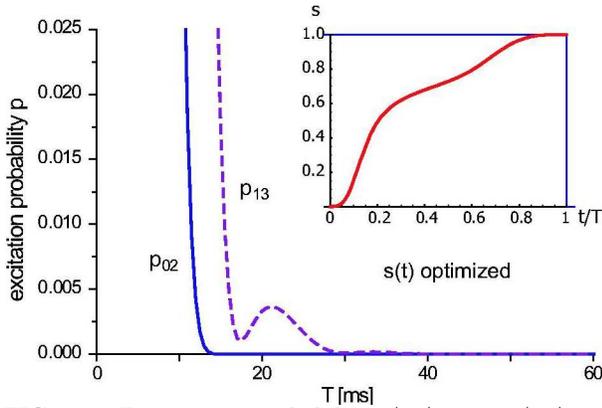} \caption{Excitation probability
 $\ket{\varphi_0}\rightarrow\ket{\varphi_2}$ and $\ket{\varphi_1}\rightarrow\ket{\varphi_3}$
 for an optimized process $s(t)$ (see inset). Note that the ordinate is scaled up by
 a factor of ten compared to fig.\,\ref{fig:ExcitationI}.}
 \label{fig:ExcitationII}
 \end{figure}

If this optimization is applied to the trapped atom
interferometer, the probability for non-adiabatic excitations can
be considerably reduced. Fig.\,\ref{fig:ExcitationII} shows the
excitation probabilities for a process speed which has been
optimized to suppress the transition
$\ket{\varphi_0}\rightarrow\ket{\varphi_1}$. With the optimized
control, the separation can be done within 30\,ms, thus reducing
the complete interferometer cycle to 60\,ms with an overall
excitation probability of less then $10^{-3}$.

These parameters suggest that an experimental realization of the
scheme is indeed feasible. It remains, of course, a difficult task
to prepare the atoms in the ground state and to detect the atoms
selectively in different vibrational states. However, this seems
achievable using Bose-Einstein condensates of low density. Another
issue is stability against gradients of magnetic stray fields. In
our case, the sensing states of the interferometer lie
$\sim6\,\mu$m apart. During a sensing time of 60\,ms, a gradient
$b_x=\frac{\partial B_x}{\partial x}$ would lead to an additional
dephasing of $\Delta\Phi \approx 2\pi\,50\cdot b_x/\GPcm$. A
suppression of stray gradients to less than 1\,mG/cm would
therefore reduce the dephasing to $\Delta\phi\leq\frac{2\pi}{20}$.

\section{Conclusion}
In conclusion, we have studied a dynamic potential interferometer
working with three-dimensionally trapped atoms. We have used a
time-dependent interaction picture to describe the quantum state
evolution and we have computed probabilities for non-adiabatic
transitions into neighboring levels. For a realistic magnetic
microtrap we find parameters which suggests an experimental
implementation in the near future. Grounding on the theoretical
results, we have developed an optimization scheme for the
reduction of vibrational excitations that is independent of the
system's parametrization. Applying the optimization to our
interferometer potential, we have found a cycle of duration
$T=60$\,ms with excitation probability less than $10^{-3}$.


\end{document}